\documentclass[aps,twocolumn,showpacs]{revtex4}
\usepackage{amssymb}
\usepackage{graphicx}
\usepackage{amsmath}
\usepackage{ulem}

\begin{document}

\title{Symmetry of the spin Hamiltonian for herbertsmithite: a spin-$\frac{1}{2}$
kagom\'{e} lattice }
\author{Oren Ofer and Amit Keren}
\affiliation{Physics Department, Technion, Israel Institute of Technology, Haifa 32000,
Israel}

\begin{abstract}
We present magnetization measurements on oriented powder of ZnCu$_{3}$(OH)$%
_{6}$Cl$_{2}$\ along and perpendicular to the orienting field. We find a
dramatic difference in the magnetization between the two directions. It is
biggest at low measurement fields $H$ or high temperatures. We show that the
difference at high temperatures must emerge from Ising-like exchange
anisotropy. This allows us to explain muon spin rotation data at $%
T\rightarrow 0$ in terms of an exotic ferromagnetic ground state.
\end{abstract}

\maketitle

The synthesis of the herbertsmithite [ZnCu$_{3}$(OH)$_{6}$Cl$_{2}$]\ \cite%
{shores} has led to a renewed interest in the frustrated spin-$1/2$\
Heisenberg model on the kagom\'{e} lattice. This system has a highly
degenerate ground state \cite{gfmRamirez} and any small perturbation to the
Hamiltonian can severely affect the ground state manifold. The perturbations
can be: exchange anisotropy \cite{starykh}, bond anisotropy \cite%
{sindzingre1,Yavorskii}, transverse field \cite{moessnerPRL,chern},
Dzyaloshinksy-Moriya Interaction\textbf{\ (}DMI) \cite%
{elhajalPRB,grohol,rigolPRB}, or longer range interactions \cite{nunez}.
Therefore, numerous theoretical directions have been taken to predict the
low-temperature behavior of the kagom\'{e} system, and some of them were
particularly applied to magnetization and other measurements of the
herbertsmithite \cite{rigolPRB}\cite{misguichSusc}.

\begin{figure}[tbp]
\includegraphics[width=8cm]{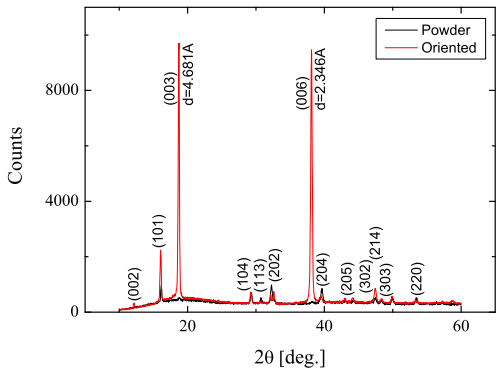}
\caption{(Color online) X-ray diffraction of powder (black) and oriented
powder (gray) from a surface perpendicular to the orienting field. }
\label{xrays}
\end{figure}

This mithite is exciting since Cu ions create a spin-$1/2$ magnetic kagom%
\'{e} layer separated by non magnetic Zn atoms from the adjacent layers. The
compound was found to be a quantum spin liquid with no broken continuous
symmetry but gapless excitations \cite%
{shlee,condmat,heltonPRL,mendelsmusr,olariu}. At high temperatures the
inverse susceptibility obeys a Curie-Weiss (CW) law, $\chi =C/(T+\theta )$,
where $C$ is the Curie constant and the CW temperature $\theta =314$~K.
Below $\sim 75$~K a sharp increase in the susceptibility occurs, deviating
from the ideal kagom\'{e} Heisenberg model \cite{NakamuraPRB95}. This upturn
was accounted for by DMI \cite{rigolPRB,zorkoESR} or anisotropy in the bonds 
\cite{sindzingre1, chern}. It was also suggested that impurities from a
Zn/Cu substitution play a significant role in the low-temperature
susceptibility \cite{olariu}\cite{fabrice,madevries}. However, free
impurities, or interacting impurities that generate an additional
ferromagnetic Curie-Weiss law \cite{rigolPRB}\cite{misguichSusc}, have been
shown not to describe this upturn completely. In fact, Rietveld refinement
of our sample showed no Zn/Cu substitution within the experimental
resolution \cite{EmilyThesis}. The sample is made by the same procedure and
group as the samples in Refs. \cite{heltonPRL} and \cite{imai}. Finally,
different local probes such as muon \cite{condmat}, O, \cite{olariu} Cu, and
Cl \cite{imai} nuclear magnetic resonance, and electron spin resonance \cite%
{zorkoESR} suggest different behavior of the susceptibility below $\sim 50$%
~K. Thus, there is still no agreement on the interactions that control the
magnetic properties of herbertsmithite.

In fact, since it is only available as a powder, the symmetries of its spin
Hamiltonian are not known.\textbf{\ }To clarify these symmetries we present
magnetization measurements on oriented powder of ZnCu$_{3}$(OH)$_{6}$Cl$_{2}$%
\ along ($\hat{z}$) and perpendicular to ($\perp $) the orienting field. The
symmetry of the interactions are probed at high temperatures where
impurities are not expected to contribute to the susceptibility even if they
do exist, and all probes roughly agree.

The orientation was done by curing ZnCu$_{3}$(OH)$_{6}$Cl$_{2}$ powder
overnight with Stycast in a field of 8 T at room temperature. The samples
were cured in a Teflon form producing a ball $6~$mm in diameter. During the
first $40$ minutes of the orientation, a shaking mechanism was applied to
the sample form. A particularly small amount of powder was used to avoid
saturating the Stycast and eliminating powder residues at the bottom of the
ball. We prepared a second \textquotedblleft test\textquotedblright\ sample
in the same manner, but this time without orientation. We refer to the
second ball as the powder sample. The mass of ZnCu$_{3}$(OH)$_{6}$Cl$_{2}$
in the ball is known only roughly and the absolute value of the molar
magnetization is not accurate. We also prepared a ball made of Stycast only.
All samples were measured in a gelatin capsule.

In Fig.~\ref{xrays} we plot the x-ray diffraction from the powder and
oriented samples. For these measurements a separate surface perpendicular to
the orienting field was prepared and used. The Bragg peak intensities are
shown in the figure. In the oriented case the (002) and (006) peaks
increased dramatically, while many of the other peaks did not. This x-ray
picture shows a high degree of orientation such that the $c$ direction is
parallel to the field. The level of orientation will be discussed further
below. 
\begin{figure}[tbp]
\includegraphics[width=7.5cm]{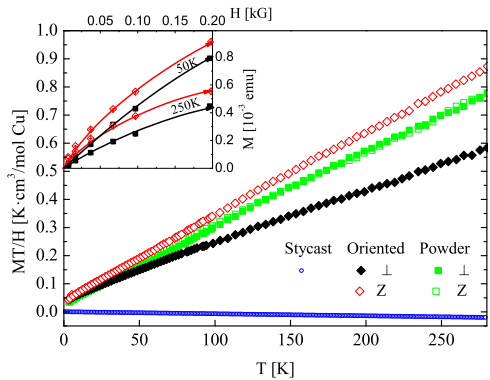}
\caption{(Color online) Normalized magnetization $M/H\equiv \protect\chi $
times the temperature versus temperature at external field of $400$~G for
oriented sample (diamonds) in two directions, powder sample (squares) in two directions as if it was oriented, and stycast sample (open circles). The inset shows the magnetization as a function of applied field for two different temperatures and directions for low fields.}
\label{chiT}
\end{figure}

DC magnetization measurements, $M$, were performed using a Cryogenic SQUID\
magnetometer in two configurations. One configuration `$z$' is when the
orienting and the applied (SQUID) fields coincide, ($H||c$). The other
configuration `$\perp $' is when the ball is rotated by $90^{\circ }$ and
thus the applied field is in the kagom\'{e} plane, ($H\perp c$). In Fig.~\ref%
{chiT} we present $MT/H$ of the two samples, powder and oriented balls%
\textbf{. } In the reset of this paper we use $\chi $ to indicate the
normalized magnetization $M/H$ (and not $\partial M/\partial H$). These
measurements were taken at $H=400$~G. The measurements are conducted as
follows: we first measured the powder sample and then the oriented sample in
both configurations. Finally, we repeated the powder measurements for a
second time, but rotated the powder ball as if it was oriented. All powder
measurements collapse into a single curve, as expected, demonstrating the
reproducibility of the measurement. The Stycast sample showed a very small
diamagnetic signal which is also depicted in Fig.~\ref{chiT}. The core
diamagnetic susceptibility of ZnCu$_{3}$(OH)$_{6}$Cl$_{2}$ is $-16.7\times
10^{-5}~$cm$^{3}$/mole \cite{Selwood}. The Van-Vleck contribution is
expected to be of the same order of magnitude, but with a positive sign \cite%
{JohnstonPRB90}. Both are much smaller than the measured $\chi $ at room
temperature of $1\times 10^{-3}~$cm$^{3}$/mole.

In Fig.~\ref{chiT} no special energy scale is found in either one of the
measurements. The only indication of an interaction between spins is the
fact that $\chi T$ for both directions and the powder decreases with
decreasing $T$. $\chi T$ of the powder is smaller than $\chi _{z}T$ and
larger than $\chi _{\perp }T$ of the oriented sample. However, a comparison
of the absolute value of $\chi $ of the powder and the oriented sample is
not accurate. We did try to have an equal amount of sample in both balls but
there is no telling how successful we were. A more relevant comparison is
between $\chi $ in the different directions of the oriented sample; $\chi
_{z}T$ increases faster than $\chi _{\perp }T$, and at room temperature $%
\chi _{z}=1.6\chi _{\perp }$. Thus the ratio between the $z$ and $\perp $
directions increases as the temperature increases.

In the inset of Fig.~\ref{chiT} we show the magnetization as a function of $%
H $ for two different temperatures and directions. The magnetization is
beginning to show signs of saturation, suggesting contribution from
ferromagnetic impurities. To check this possibility we present in Fig.~\ref%
{Arrott} an Arrott plot~\cite{ArrottPR57}. This plot takes advantage of the high field data.
At a ferromagnetic transition, $M^{2}(T_{c})$ is expected to be a linear
function of $H/M$. We found no evidence for such linear behavior. In fact, $%
M^{2}(T)$ is independent of $H/M$ near the origin as expected when the
ferromagnetic critical temperature is lower than the available temperature.
This indicates the lack of ferromagnetic impurities in our measurements.

\begin{figure}[tbp]
\includegraphics[width=7.5cm]{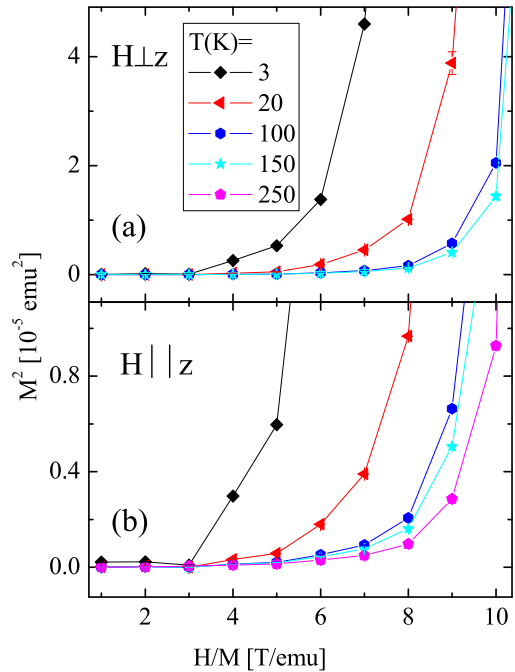}
\caption{ (Color online) An Arrott plot showing $M^{2}$ versus $H/M$ at
various temperatures in the perpendicular direction (a) and parallel
direction (b). }
\label{Arrott}
\end{figure}

\begin{figure}[bp]
\includegraphics[width=7cm]{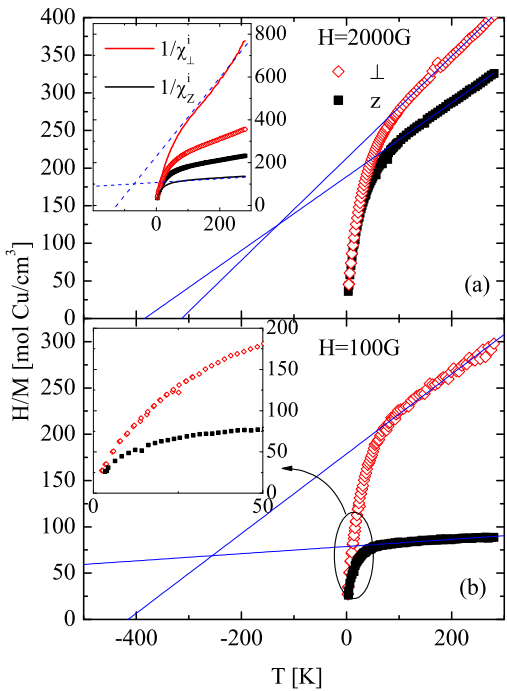}
\caption{(Color online) Inverse normalized magnetization $H/M\equiv \protect%
\chi ^{-1}$ versus temperature at $H=2$ kG (a) and at $H=100$ G (b). The
solid lines are linear fits to the high temperature ($>150$~K) data. (a)
inset displays the inverse of the normalized magnetizations $1/\protect\chi %
_{z}$ (black squares) and $1/\protect\chi _{\perp}$ (gray diamonds), and the inverse intrinsic normalized
magnetization $1/\protect\chi _{z}^{i}$ (black line) and $1/\protect\chi _{\perp }^{i}$ (gray line)
obtained from Eq.~\protect\ref{inverting}. The dashed lines demonstrate that 
$\protect\theta _{Z}<\protect\theta _{Z}^{i}$ and $\protect\theta _{\perp }>%
\protect\theta _{\perp }^{i}$. In the inset of (b) we plot the
low-temperature behavior of $\protect\chi ^{-1}$ at $100$~G.}
\label{ichi}
\end{figure}

In Fig.~\ref{ichi} we plot $\chi ^{-1}$ versus temperature for two fields, $%
2000$ and $100$~G and for the two orientations. In the inset of Fig.~\ref%
{ichi}(b) we plot the $\chi ^{-1}$ at low temperatures ($T<50$~K); clearly, $%
\chi _{z}$ linearizes at $T\sim 30$~K whereas $\chi ^{\perp }$ linearizes at
a much higher temperature ($T\sim 100$~K). $\theta $, and $C$ in arbitrary
units are extracted from a linear fit of the high-temperature ($150<T<280$%
~K) data to $\chi _{\perp ,z}^{-1}=(T+\theta _{\perp ,z})/C_{\perp ,z}$. The
fits are shown by the solid line.

\begin{figure}[tbp]
\includegraphics[width=8cm]{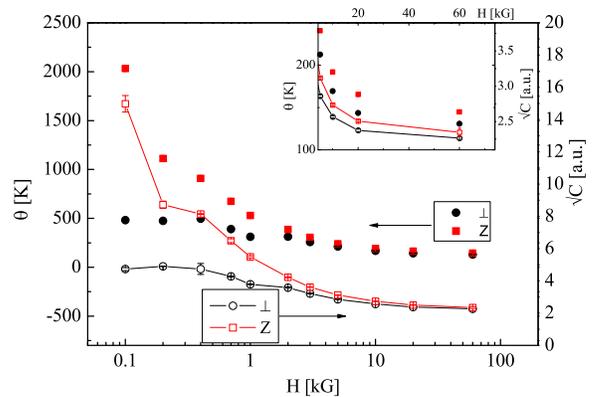}
\caption{(Color online) The Currie-Wiess temperatures (filled symbols) and square root of the
Currie constant (open symbols) of the oriented sample perpendicular to (black) and in the kagom\'{e}
plane (gray). The inset show a zoom on the high field data. }
\label{tcw}
\end{figure}

In Fig.~\ref{tcw} we plot $\theta _{\perp ,z}$, and $\sqrt{C_{\perp ,z}}$
which is proportional to the $g_{\perp ,z}$ factor (if the sample was fully
oriented) versus the applied field. $\theta _{\perp }$ increases slowly with
decreasing applied field and saturates below $400$~G. On the other hand, $%
\theta _{z}$ increases rapidly below $2$~kG. The Curie constant has a
similar behavior. The powder average of $\theta _{\perp ,z}$ at low fields
does not reconcile with $\theta \sim 300$~K measured in a powder and there
must be some extrinsic contribution to the normalized magnetization in the
partially aligned samples at low fields. However, we have no evidence that
this contribution is due to impurities.

In contrast, at high fields, $H>2$~kG, $\theta $ of the two directions is
hardly distinguishable and on the order of the powder value. In addition,
useful information can be extracted from the CW temperature only if it is
obtained by measurements at $T\gtrsim \theta $. Therefore, we concentrate on
the results obtained by $H\geq 2$~kG, as shown in the inset of Fig.~\ref{tcw}%
. At $2$~kG the ratio of $\sqrt{C_{z}/C_{\bot }}=1.179(6)$ and $\theta
_{z}/\theta _{\bot }=1.23(1)$.

In order to convert the measured $\chi $ presented above to the intrinsic
normalized magnetization $\chi ^{i}$ in different directions, it is
important to estimate the level of orientation. This can be done using the
x-ray data. The ratio of the x-ray intensity ($I$) from the powder $%
R=I(00h)/I(kk0)$ represents the signal intensity ratio between the two kinds
of plane. Let's assume that there are $N$ grains composed of two sets, $%
\alpha N$ that can orient perfectly with the field, and $(1-\alpha )N$ that
are not effected by the field at all since they are made of a few
crystalline, for example. We further define $\beta $ as the probability that
a particular plane will contribute to the scattering intensity in a powder.
After orientation the x-ray intensity ratio between the same planes would be 
$R^\prime =[\alpha R+\beta (1-\alpha )R]/\beta (1-\alpha )$. We can estimate $%
\beta $ from the width of the peaks which is $0.2^{\circ }$ out of $180$,
thus $\beta \sim 0.001$. Using $R$ and $R\prime $ of the $(006)$ and $(220)$
we find $\alpha =0.25$. This level of orientation is in agreement with Imai 
\textit{et al.}\cite{imai}.

In an oriented sample we expect 
\begin{equation}
\chi _{z,\perp }=(1-\alpha )\left( \frac{1}{3}\chi _{z}^{i}+\frac{2}{3}\chi
_{\perp }^{i}\right) +\alpha \chi _{z,\perp }^{i}.  \label{inverting}
\end{equation}%
This relation could be inverted to produce $\chi _{z,\perp }^{i}$. In the
inset of Fig.~\ref{ichi}(a) we present both $1/\chi _{z,\perp }$ and $1/\chi
_{z,\perp }^{i}$ for the normalized magnetization data taken at $H=2$~kG.
New intrinsic CW temperatures $\theta _{z,\perp }^{i}$ could be obtained
from $1/\chi _{z,\perp }^{i}$ as demonstrated by the dashed lines. $\theta
_{z,\perp }^{i}$ represent the CW temperature as if the sample was fully
oriented. Although $\alpha $ is just an estimate of the level of
orientation, the important point is that $\theta _{z}^{i}>\theta _{z}$ and $%
\theta _{\perp }^{i}<\theta _{\perp }$.

We now turn to discuss the possible origin of the $\chi $ anisotropy in
terms of superexchange anisotropy and DMI. The DMI Hamiltonian is given by, 
\begin{equation}
\mathcal{H}=\sum_{<i,j>}J\mathbf{S}_{i}\cdot \mathbf{S}_{j}+\mathbf{D}%
_{ij}\cdot \left( \mathbf{S}_{i}\times \mathbf{S}_{j}\right)
\label{hamiltonian}
\end{equation}%
where $\mathbf{D}_{ij}$ is a vector assigned to each bond. In the mean field
approximation ($\mathbf{S}_{j}\rightarrow \mathbf{M}/g\mu _{B}$) this
Hamiltonian is written as 
\begin{equation*}
\mathcal{H}=g\mu _{B}\sum_{i}\mathbf{S}_{i}\cdot \mathbf{H}^{eff}
\end{equation*}%
where%
\begin{equation}
\mathbf{H}^{eff}=\frac{Z}{(g\mu _{B})^{2}}\left( J\mathbf{M}+\mathbf{D}%
\times \mathbf{M}\right) +\mathbf{H}
\end{equation}%
$\mathbf{D=}(1/Z)\sum_{j}\mathbf{D}_{ij}$, and $Z$ is the number of
neighbors. Special attention must be taken for the convention of the $ij$
bond direction since it sets the direction of $\mathbf{D}_{ij}$ \cite%
{rigolPRB}. The magnetization satisfy the equation 
\begin{equation}
\mathbf{M}=\frac{C}{T}\left( \frac{Z}{(g\mu _{B})^{2}}\left( J\mathbf{M}+%
\mathbf{D}\times \mathbf{M}\right) +\mathbf{H}\right)  \label{mag1}
\end{equation}%
where $C=(g\mu _{B})^{2}S(S+1)/(3k_{B})$ is the Curie constant. Up to first
order in $\mathbf{D}$ 
\begin{equation}
\mathbf{M}=\frac{C}{(T-\theta_{cw})}\left( I+\frac{1}{T-\theta _{cw}}%
\mathbf{A}\right) \mathbf{H.}
\end{equation}%
where $\theta _{cw}=CZJ/(g\mu _{B})^{2}$ and%
\begin{equation}
\mathbf{A}=\frac{CZ}{(g\mu _{B})^{2}}\left( 
\begin{array}{ccc}
0 & -D_{z} & D_{y} \\ 
D_{z} & 0 & -D_{x} \\ 
-D_{y} & D_{x} & 0%
\end{array}%
\right) ~.
\end{equation}%
In particular 
\begin{equation}
M_{z,\perp }=\frac{C}{(T-\theta _{cw})}H_{z,\perp }
\end{equation}%
Therefore, $\mathbf{D}_{ij}$ does not contribute to the CW law.

In contrast, the superexchange anisotropy Hamiltonian is given by 
\begin{equation}
\mathcal{H}=\sum_{<i,j>}J_{z}S_{i}^{z}\cdot S_{j}^{z}+J_{\perp }\mathbf{S}%
_{i}^{\perp }\cdot \mathbf{S}_{j}^{\perp }.
\end{equation}
In this case, if the sample was perfectly oriented, we would have $\theta
_{z,\perp }=J_{z,\perp }/k_{B}$. Since our sample is not perfectly
oriented, our high-temperature high field linear fits of $\chi _{\perp
,z}^{-1}$ measures a lower bound on $J_{z}$ and an upper bound on $J_{\perp }
$.

The lower bound on $J_{z}$ is larger than the upper bound on $J_{\perp }$.
Despite the fact that measurements of $\chi _{z}$ and $\chi _{\perp }$ are
contaminated with $\chi _{\perp }^{i}$ and $\chi _{z}^{i}$ respectively, as
indicated by Eq.~\ref{inverting}, the conclusion $J_{z}>J_{\perp }$ is
unavoidable. It is robust even against possible core and Van-Vleck
corrections. Thus herbertsmithite has an Ising-like exchange anisotropy.
This, however, is not the end of the story. If $J_{z}>J_{\perp }$, we would
expect $\chi _{z}<\chi _{\perp }$, in contrast to observation. Therefore, to
explain the high $\chi $ in the $z$ direction we must invoke an anisotropic $%
g$ factor as well.

In the classical ground state of antiferromagnets on the kagom\'{e} lattice
with exchange anisotropy, the spins are coplanar and two angles between
spins $\varphi $ on each triangle obey $\cos \varphi =-J_{z}/(J_{z}+J_{\perp
})$. The third angle completes the circle. This condition maintains the
ground state macroscopic degeneracy. Nevertheless, unlike in the Heisenberg
case, there is a critical temperature $T_{c}$ below which an exotic
ferromagnetic order exists with finite total magnetization, but no
sublattice long-range order \cite{KurodaJPSJ95}. Upon cooling through $T_{c}$
the magnetization increases abruptly and continuously down to $T\rightarrow
0 $ where it saturates \cite{TanakaJPCM07}. In zero field, domains can be
formed, but a small applied magnetic field will stabilize the moment. The
powder average of the moment projection on the field direction is given by
the value 
\begin{equation}
\left\langle \mathbf{M\cdot }\widehat{\mathbf{H}}\right\rangle =\frac{\mu
_{B}}{6}(1+2\cos \varphi )  \label{Moment}
\end{equation}%
per spin.

We believe that this ferromagnetic order contributes to the observed $\chi $
at $T\rightarrow 0$ by transverse field (TF) muon spin rotation ($\mu $SR)
experiment \cite{condmat}. In $\mu $SR, impurities, if they exist, are
expected to contribute to the muon line width while most of the sample
contributes to the line shift. In what follows we examine what part of the $%
\mu $SR shift can be explained by exchange anisotropy only. A complete
understanding will of course require taking DMI interaction into account as
well.

\begin{figure}[tbp]
\begin{center}
\includegraphics[width=9cm]{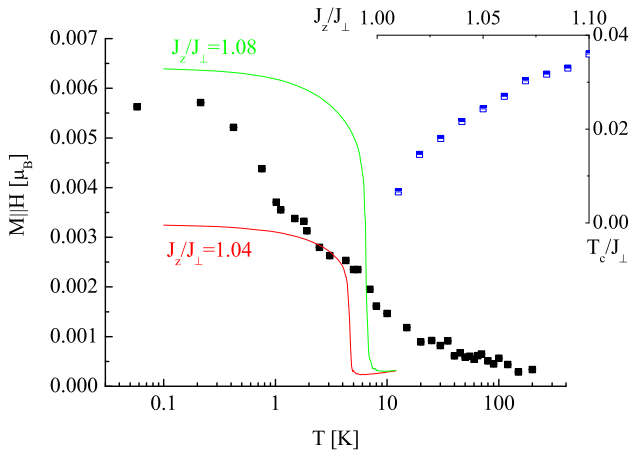}
\end{center}
\caption{(Color online) A plot of the magnetization detected by muon spin
rotation versus temperature (black squares), and simulation data for antiferromagnetic kagom%
\'{e} lattice with Ising-like exchange anisotropy as in Ref.~\protect\cite%
{TanakaJPCM07} (gray lines). In the inset the normalized critical temperature versus the
exchange anisotropy is shown.}
\label{musr}
\end{figure}

The $\mu $SR measurements were done at a field of $H=2$~kG and the shift $K$
in the muon rotation frequency as a function of temperature was measured.
This shift is a consequence of the sample magnetization; therefore, $K$ is
expected to be proportional to normalized magnetization. The high
temperature data are used to calibrate the proportionality constant between $%
K$ and $\chi $ \cite{condmat}. The data are reproduced in Fig. \ref{musr}. $%
\chi $ increases sharply with decreasing temperatures between $\sim 10$~K
and $\sim 1$~K and saturates below $T\sim 200~$mK at a value of $\chi
=15.7(5)\times 10^{-3}$ cm$^{3}$/mol Cu. This $\chi $ mounts to an average
moment of $0.006\mu _{B}$ per Cu, in the direction of the applied $2$~kG
field. Solving Eq.~\ref{Moment} for the anisotropies gives $J_{z}/J_{\bot
}=1.06$. In Fig. \ref{musr} we present simulations described in Ref.~\cite%
{KurodaJPSJ95}, for $J_{z}/J_{\bot }=1.04$ and $J_{z}/J_{\bot }=1.08$
showing similar behaviour as the experiment. For this type of exchange
anisotropy the expected $T_{c}/J_{\bot }=0.03$ as shown in the inset of Fig. %
\ref{musr} also taken from Ref.~\cite{KurodaJPSJ95}. For $J_{\bot }\simeq
200 $~K we expect $T_{c}=6$~K. This temperature is at the center of the
sharp rise of $\chi $. Thus we see that both the saturation and the increase
of $\chi $ detected by $\mu $SR could be qualitatively explained by exchange
anisotropy. 

To summarize, our measurements in ZnCu$_{3}$(OH)$_{6}$Cl$_{2}$ reveal an
anisotropic intrinsic spin magnetization with $\chi _{z}^{i}>\chi _{\perp
}^{i}$ possibly due to anisotropic $g$ factor. At fields above $2$~kG a CW
temperature can be consistently determined in two different directions. By
mean-field approximations we were able to show that this phenomenon can be
explained only by anisotropic super-exchange constants where $J_{z}>J_{\bot
} $. This anisotropy can explain the main features of the susceptibility
determined by $\mu $SR.

We are grateful to E. A. Nytko and D. G. Nocera for providing us with the
sample, and to S. Tanaka, S. Miyashita, and N. Kawashima for providing us
with the simulation results shown in Fig. \ref{musr}. We acknowledge helpful
discussions with Young S. Lee, Rajiv Singh, Marcos Rigol, and John Chalker.
We also would like to thank the Israel - U. S. Binational Science Foundation
for supporting this research.

\end{document}